\documentstyle[12pt,epsfig]{elsart}
\def\Journal#1#2#3#4{{#1} {\bf #2}, #3 (#4)}

\def\NPB{{\em Nucl. Phys.} {\bf B}}

\def\PLB{{\em Phys. Lett.}  {\bf B}}
\def\PRL{\em Phys. Rev. Lett.}
\def\PRD{{\em Phys. Rev.} {\bf D}}

\def\PRP{{\em Phys. Rep. }}

\def\ZPC{{\em Z. Phys.} C}

\def\MPL{{\em Mod. Phys. Lett. }}

\def\PNPP{\em Prog. Nucl. Part. Phys.}

\def\PTPS{\em Prog. Theo. Phys. Suppl.}

\def\ra{\rightarrow}
\def\be{\begin{equation}}
\def\ee{\end{equation}}

\newcommand{\ls}{\mbox{$\stackrel{<}{\sim}$ }}
\newcommand{\expe}{experiment}

\newcommand{\osc}{oscillations}
\newcommand{\expes}{experiments}

\newcommand{\br}{branching ratio}
\newcommand{\bb}{double beta decay}

\newcommand{\nbb}{neutrinoless double beta decay}
\newcommand{\majo}{Majorana}

\newcommand{\delm}{\mbox{$\Delta m^2$} }

\newcommand{\nel}{\mbox{$\nu_e$}}
\newcommand{\nmu}{\mbox{$\nu_\mu$}}

\newcommand{\neu}{neutrino}
\newcommand{\neus}{neutrinos}

\newcommand{\ema}{\mbox{$\langle m_{ee} \rangle$ }}

\newcommand{\nhls}{neutral heavy leptons}
\newcommand{\ba}{\begin{array}{c}}
\newcommand{\baz}{\begin{array}{cc}}
\newcommand{\bad}{\begin{array}{ccc}}
\newcommand{\bea}{\begin{equation} \begin{array}{c}}
\newcommand{\eea}{ \end{array} \end{equation}}
\newcommand{\ea}{\end{array}}
\newcommand{\D}{\displaystyle}

\newcommand{\mab}{\mbox{$\langle m_{\alpha \beta} \rangle $}}
\newcommand{\mmm}{\mbox{$\langle m_{\mu \mu} \rangle $}}
\newcommand{\mtt}{\mbox{$\langle m_{\tau \tau} \rangle $}}

\newcommand{\mee}{\mbox{$\langle m_{ee} \rangle $}}

\newcommand{\mem}{\mbox{$\langle m_{e \mu} \rangle$}}
\begin{document}
\begin{frontmatter}
\title{New limits on effective Majorana neutrino masses from rare kaon decays}
\author{K. Zuber}
\address{Lehrstuhl f\"ur Experimentelle Physik IV, Universit\"at Dortmund,
Otto-Hahn Str.4, 44221 Dortmund, Germany}
\begin{abstract}
The rare kaon decay $K^+ \ra \pi^- \mu^+ \mu^+$ is violating lepton number by two units and is investigated
within the context of Majorana neutrino masses. Using
a new upper bound from E865 an upper limit on the effective Majorana mass term
$\mmm \ls 500$ GeV could be obtained. This is improving
existing bounds by about one order of magnitude. Implications on heavy neutrinos are discussed as well
as future possibilities to improve limits on this as well as other elements of the Majorana mass matrix
are investigated. 
\end{abstract}
\end{frontmatter}

\section{Introduction}
Over the last years growing evidence arose for a non-vanishing \neu{} rest mass.
The observations of missing solar \neus, a deficit in upward going atmospheric \nmu{} and
the LSND accelerator \expe{} results all can be explained within the context of \neu{} \osc. For recent reviews
see
\cite{zub98,bil98}.
On the other
hand oscillation
\expes{} are no absolute
mass measurements, depending on \delm{} = $m_2^2 - m_1^2$, where $m_{1,2}$ are the two mass eigenvalues. Therefore
several \neu{} mass models exist 
to describe the observed effects. 
Beside
that also the fundamental character of the \neu , whether being a Dirac or Majorana particle, is 
still unknown. The most promising channel to probe this property for \nel{} is \nbb. The measured
quantity \mee{} is called effective \majo{} \neu{} mass and given by
\be
\label{eq:1}
\mee \! \! = | \sum U_{ei}^2 m_i \eta^{\rm CP}_i |
\ee
where $m_i$ are the mass eigenvalues, $\eta^{\rm CP}_i = \pm 1$ are the relative CP-phases and $U_{ei}$ are the
mixing matrix elements connecting weak eigenstates with mass eigenstates. The current 
experimental upper bound on \mee{} is around 0.2 eV \cite{bau99}. But this quantity is only one element
of a
general $3 \times 3$ matrix of effective \majo{} \neu{} masses given in the form
\bea
\label{eq:2}
\mab  = 
\left| \sum U_{\alpha i} U_{\beta i} m_i \eta^{\rm CP}_i  \right|
\mbox{ with } 
\alpha, \, \beta = e , \, \mu , \, \tau  . 
\eea
The limits for the other matrix elements are rather poor compared with \bb{}. Limits on \mem{}
arise from muon-positron conversion on titanium coming from the experimental bound of \cite{kau98}
\be
\label{eq:3}
\frac{\D \Gamma ({\rm Ti} + \mu^- \to  {\rm Ca}^{GS} + e^+)}
{\D \Gamma ({\rm Ti} + \mu^- \ra Sc + \nmu)} 
< 1.7 \cdot 10^{-12} \quad (90 \% \mbox{CL}) 
\ee
which can be converted in a new limit of \mem $<$ 17 (82) MeV depending on whether
the proton pairs in the final state are in a spin singlet or triplet state and allowing correction
factors of the order one for the difference in Ti and S as given in \cite{doi85}. 
Recently an improved
limit on the element \mmm  \ls $10^4$ GeV was given by investigating trimuon production in \neu{} - nucleon
scattering \cite{frz98}.
The first full matrix of limits on \mab{}, including for the first time matrix elements containing the $\tau$ -
sector of Eq. \ref{eq:2} are
 given in \cite{frz98a}, using HERA charged current results with associated dimuon production. 
All limits obtained for the matrix elements are of the order $10^4$ GeV or slightly below.
Indirect bounds coming from FCNC processes like $\mu \ra e \gamma, \tau \ra \mu \gamma$ could 
also account for severe limits. On the other hand note, that these processes depend on
$m_{e\mu}= \sqrt{\sum U_{ei} U_{\mu i} m_i^2}$ (in case of $\mu \ra e \gamma$) 
while the ones mentioned before typically depend on \mab$^2$. Therefore without specifying a mixing
and mass scheme, the quantities are rather difficult to compare. A discussion of such models within this context
and using oscillation data can be found in \cite{wer00}. The
same argument holds for combining \mee{} and \mem{} to determine \mmm. Therefore any experimentally obtained 
limit is very useful.\\
A further interesting topic within this context is the production of neutral heavy leptons and direct production 
of \majo{} \neus{} heavier than 100 GeV, the last
has been studied for various collider types \cite{col99}. The current limits on \nhls{} are coming from LEP and are 
39.5 GeV (stable) \cite{abr97} and 76.0 GeV (for an unstable \majo{} \neu{} coupling to muons) \cite{opal00}. Also
mixing of
such 
heavy particles with the light \neus{} will be
restricted by the limits given for \mab . 
The process of a lepton-number violating rare
kaon decay discussed in this paper will allow to obtain a new limit on \mmm .

\section{The decay $K^+ \ra \pi^- \mu^+ \mu^+$}
A further possibility to probe \mmm is the rare kaon decay 
\be
K^+ \ra \pi^- \mu^+ \mu^+ \quad .
\ee
This process is violating lepton number by two units. The measured quantity \mmm{} is
given by
\be
\mmm \! \! = | \sum U_{\mu i}^2 m_i \eta^{\rm CP}_i | \quad .
\ee
The lowest order Feynman diagrams are shown in Fig. 1. 
The amplitude $A_1$ is given for the tree diagram by \cite{lit92}
\be
A_1 = 2 G_F^2 f_K f_\pi (V_{ud}V_{us})^{\ast} \sum_i (U_{li}U_{l'i})^{\ast} p_{K,\alpha} p_{\pi,\beta} [
L_i^{\alpha \beta} (p_l,p_{l'}) - \delta_{ll'} L_i^{\alpha \beta}(p_{l'},p_{l}) ]
\ee
where
\be
L_i^{\alpha \beta} (p_l,p_{l'}) = m_{\nu_i} (q^2 - m_{\nu_i}^2)^{-1} \bar{v}(p_l) \gamma^\alpha \gamma^\beta P_R
v^c (p_{l'})
\ee 
and q corresponds to the four-momentum of the virtual $\nu_i$ and $P_R = (1 + \gamma_5)/2$.
The box diagram cannot be calculated easily because the hadronic matrix element
\be
\int d^4x d^4y e^{i(p_d-p_u)y}e^{i(p_{\bar{s}} - p_{\bar{u}})x} \langle \pi^- \mid [\bar{d_L} (y) \gamma_\beta
u_L(y)][\bar{s_L}(x) \gamma_\alpha u_L(x)] \mid K^+ \rangle 
\ee
is not directly related to measured quantities like $\langle 0 \mid \bar{s_L} \gamma_\alpha u_L \mid K^+ \rangle$
and $\langle \pi^- \mid \bar{d_L}\gamma_\beta u_L \mid 0 \rangle$ 
as in the tree graph.
The tree diagram is dominating the total decay rate
and the $m_{\nu}$ dependence is coming basically from $L^{\alpha \beta}$.
Detailed calculations can be found in \cite{lit92,hal76}. Because we are far away from the expected
rate,
the uncertainties in the matrix element can be neglected. 
A first extraction of a branching
ratio for this process was done in \cite{lit92} reexamining the data from \cite{cha68}. They obtained a \br{} of
\be
\frac{\D \Gamma (K^+\to \pi^- \mu^+ \mu^+)}{\D \Gamma (K^+ \to {\rm all})} 
< 1.5 \cdot 10^{-4} \quad (90 \% \mbox{CL})
\ee
Using this value with the theoretical calculations of \cite{doi85} a limit of \mmm $< 1.1 \cdot 10^5$ GeV could be
deduced \cite{nis99}.
The processes discussed in \cite{frz98a} were able to improve that number down to be less than $4 \cdot 10^3$ GeV. 
In the meantime new sensitive kaon experiments are online and using the E865 \expe at BNL
a new upper limit on the \br{} of 
\be
\frac{\D \Gamma (K^+ \to \pi^-\mu^+ \mu^+)}{\D \Gamma (K^+ \to {\rm all})} 
< 3 \cdot 10^{-9} \quad (90 \% \mbox{CL})
\ee
could be obtained \cite{ma99}, an improvement by a factor 50000. Because the \br{} is $\propto \mmm^2$
this can be converted in a limit on \mmm \ls 500 GeV, a factor of eight better than the existing
limits and three orders of magnitude better in this particular decay channel.

\section{Results and Discussion}
The obtained upper limit on \mmm{} of 500 GeV restricts regions of heavy \majo{} \neus{}
having a mixture U$_{\mu H}$ with \nmu. No direct bound exists for neutrino masses heavier than 90 GeV. This is
illustrated in Fig. 2.
Further improvements to this bound could come from even more sensitive searches for
this rare kaon decay. An improvement by a factor of about 10 on \mmm{} implying an improvement on the \br{} 
limit by two orders of magnitude would bring the number in overlap with LEP searches. 
New experiments like E949 at BNL and CKM (E905) at Fermilab \cite{ket00} or a muon collider could improve on that
significantly. 
Furthermore the decay of charmed mesons could be considered as
well. Among the Cabibbo angle favoured modes are $D^+ \ra K^- \mu^+ \mu^
+, D_S^+ \ra
K^- \mu^+ \mu^+$ or $D_S^+ \ra
\pi^- \mu^+ \mu^+$. The existing limits on the \br{} for these processes are $3.2 \cdot 10^{-4}, 5.9 \cdot
10^{-4}$ and $4.3 \cdot 10^{-4}$
respectively \cite{kod95}. While being competitive with the old bound for the kaon decay discussed, the
new kaon \br{} limit is now five orders of magnitude better.
Therefore, to obtain new information on \mmm{} from D-decays, analyses of new data sets have to be done.\\
To improve significantly towards lighter \neu{} masses (\mmm \ls 1 GeV) 
one might consider other processes. 
The close analogon to \bb{} and therefore a measurement of \mmm{} using nuclear scales would be
muon capture by nuclei with a $\mu^+$ in
the final state
as discussed in \cite{mis94} . No such \expe{} was performed yet.
The ratio with respect to muon capture can be given for $^{44}$Ti and light neutrino exchange as
\be
\Gamma = \frac{\Gamma (\mu^- + {\rm Ti} \ra \mu^+ + {\rm Ca})}{\Gamma (\mu^- + {\rm Ti} \ra \nmu + {\rm Sc})}
\simeq 5
\cdot 10^{-24} (\frac{\mmm}{250
keV})^2 \quad .
\ee 
Assuming that a \br{} of the order of muon-positron
conversion (Eq.\ref{eq:3}) can be obtained, a bound on \mmm \ls 150 GeV results. Unfortunately
this is only a slight
improvement on the bound obtained here. Furthermore assuming heavy neutrino exchange for the $\mu^+$ capture would
result in a rate another four orders of magnitude lower than for light neutrino exchange.
Improvements on the $\tau$ - sector of matrix elements of Eq. (\ref{eq:2}), especially \mtt, could
be done by
a search for rare B-decays. Limits on the \br{} for decays $B^+ \ra K^- \mu^+ \mu^+, B^+ \ra \pi^- 
\mu^+ \mu^+$ of less than $9.1 \cdot 10^{-3}$ exist \cite{wei90}, however nobody looked into the decays $B^+ \ra
K^- \tau^+ 
\tau^+$ or $B^+ \ra \pi^- \tau^+ \tau^+$. With the new B-factories such a search might be possible
at a level of producing limits on \mtt{} competitive with the ones given in \cite{frz98a}.

\section{Conclusion}
Whether \neus{} are Dirac or \majo{} particles is still an open question. For several flavours
an effective Majorana mass matrix can be assumed, whose best explored element is \ema due
to \nbb{} searches. An improved limit on $\langle m_{e \mu} \rangle$ is given here.
An investigation especially on the matrix element \mmm{} was performed. Using
new bounds on the \br{} of $K^+ \ra \pi^- \mu^+ \mu^+$ obtained by
E865 a new upper limit on \mmm{} of less than 500 GeV was obtained, improving existing
bounds by roughly one order of magnitude. Informations on the admixture of heavy \majo{}
\neus{} with muons are obtained. 
Suggestions for further improvements are given as well
as the proposal to consider a search for rare B - decays like $B^+ \ra \pi^- \tau^+ \tau^+$ to improve on the 
$\tau$ - sector of \mab.
\section{Acknowledgements}
I would like to thank H. Ma for providing me the new E865 limit and W. Rodejohann
for useful discussions.

\newpage
\begin{center}
\begin{figure}
\begin{tabular}{cc}
\mbox{\epsfig{file=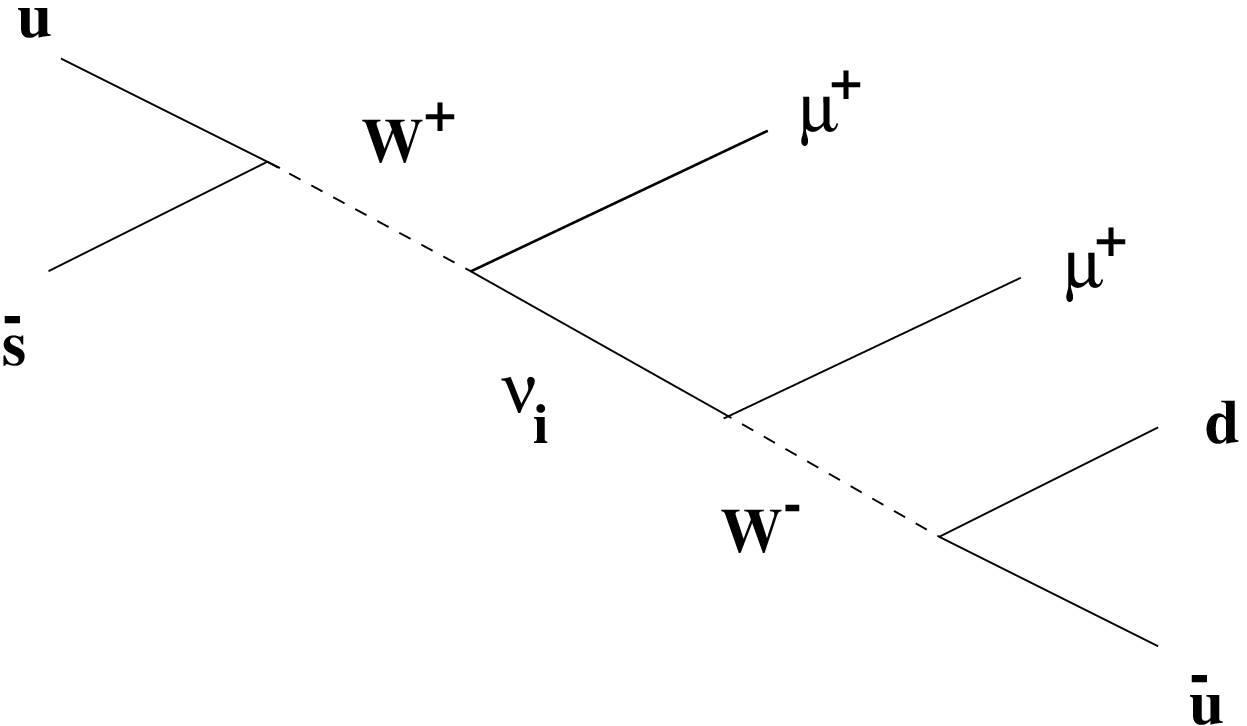,height=5cm,width=7cm}} &
\mbox{\epsfig{file=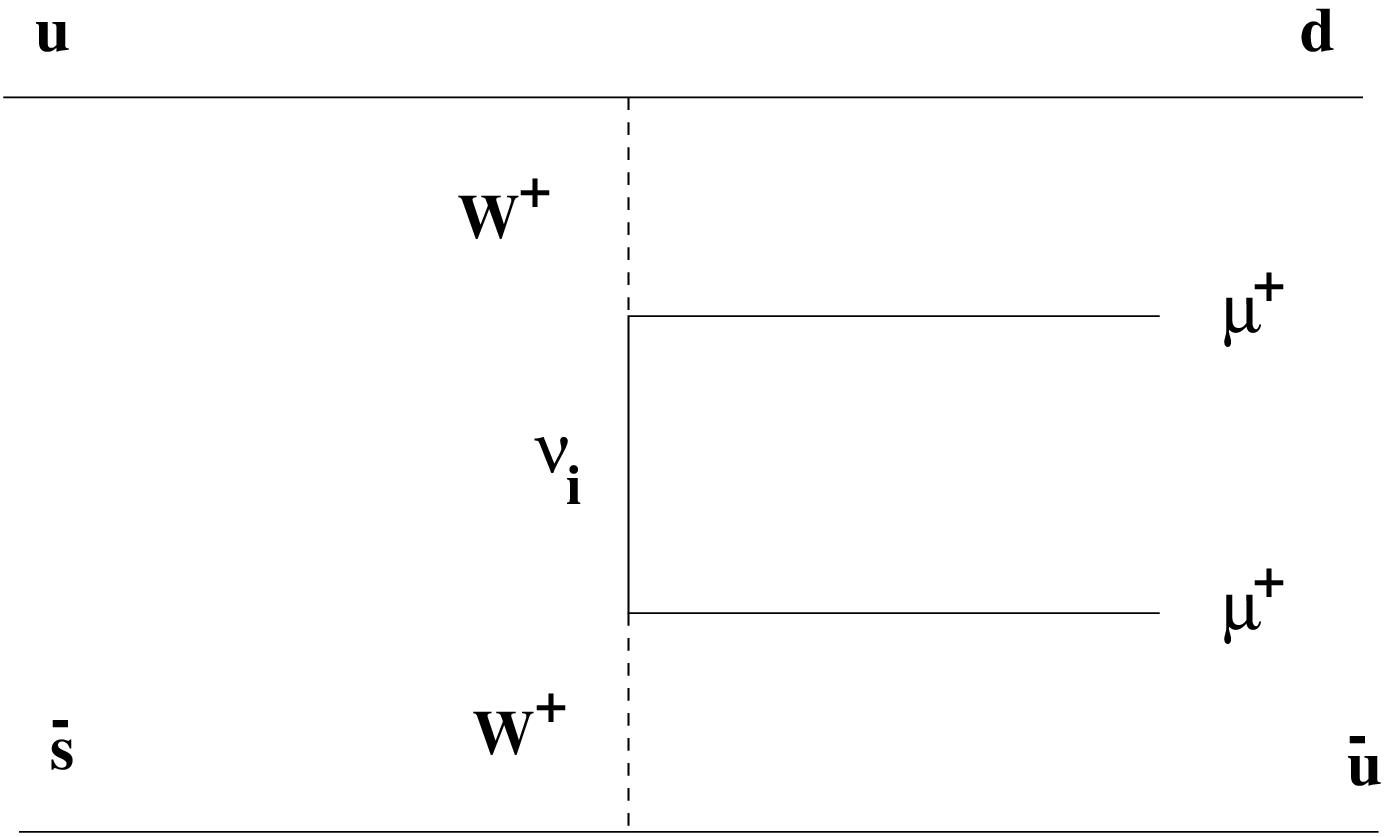,height=5cm,width=7cm}}
\end{tabular}
\vspace{2mm}
\label{pic:feyn}
\caption{Feynman diagrams in lowest order contributing to the rare kaon decay 
$K^+ \ra \pi^- \mu^+ \mu^+$. Shown are the tree (left) and box diagram (right).}
\end{figure}
\end{center}
\begin{center}
\begin{figure}
\epsfig{file=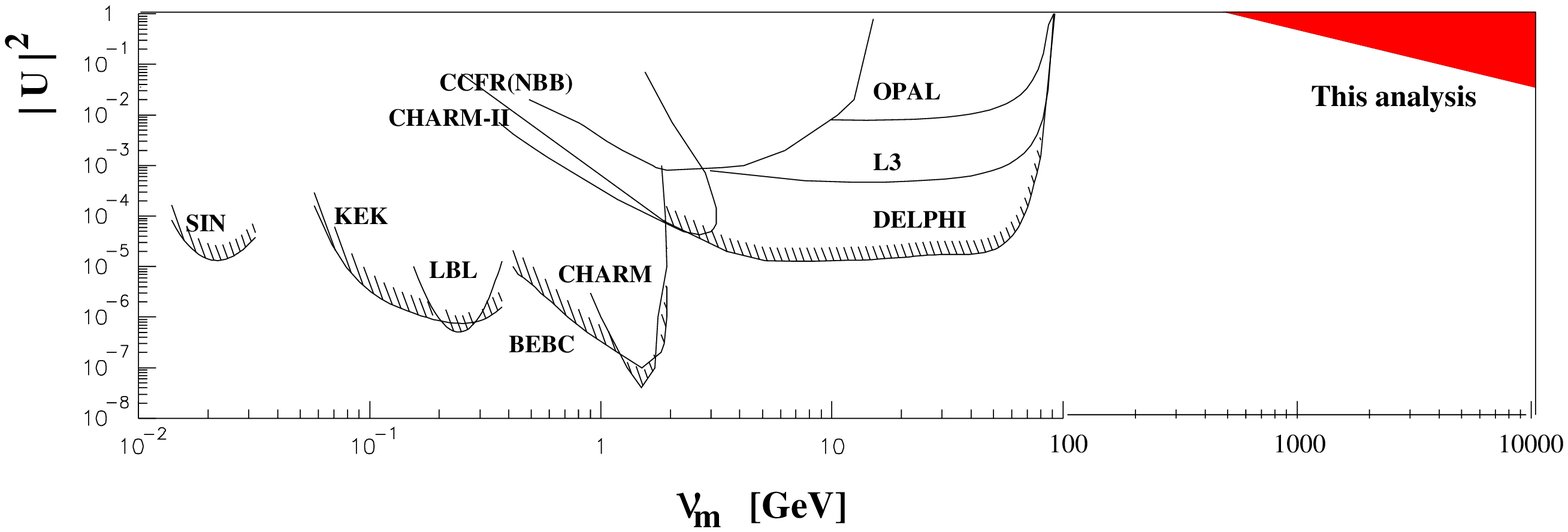,height=7cm,width=14cm}
\vspace{2mm} 
\label{pic:limits}
\caption{95 \% CL Limits on the mixing matrix element $\mid U^2 \mid$ of a heavy neutrino to a light one.
The compilation of limits below 90 GeV are taken from \protect \cite{abr97}. No direct limit 
exist for the mixing matrix element $\mid U_{\mu H} \mid ^2$ above 90 GeV. The region which can
be excluded by the analysis of rare kaon decays is shown in the upper right.} 
\end{figure}
\end{center}


\begin{thebibliography}{99}
\bibitem{zub98} K. Zuber, \Journal{\PRP}{305}{295}{1998}
\bibitem{bil98} S. M. Bilenky, C. Giunti, W. Grimus, \Journal{\PNPP}{43}{1}{1999}
\bibitem{bau99} L. Baudis et al., \Journal{\PRL}{83}{411}{1999}
\bibitem{kau98} J. Kaulard et al., \Journal{\PLB}{422}{334}{1998}
\bibitem{doi85} M. Doi, T. Kotani, E. Takasugi, \Journal{\PTPS}{83}{1}{1985} 
\bibitem{frz98} M. Flanz, W. Rodejohann, K. Zuber, preprint hep-ph/9907203
\bibitem{frz98a} M. Flanz, W. Rodejohann, K. Zuber, \Journal{\PLB}{473}{324}{2000}
\bibitem{wer00} W. Rodejohann, preprint hep-ph/0003149
\bibitem{col99}W. Buchm\"uller, C. Greub, \Journal{\NPB} 
{363}{345}{1988}, \Journal{\NPB}{381}{109}{1992};
J. Gluza, M. Zralek, \Journal{\PRD}{48}{5093}{1993}, A. Hoefer, L. M. Sehgal, \Journal{\PRD} 
{54}{1944}{1996}, G. Cvetic, C. S. Kim, \Journal{\PLB}{461}{248}{1999}
\bibitem{abr97} P. Abreu et al., Delphi-coll., \Journal{\ZPC}{74}{57}{1997}
\bibitem{opal00} G. Abbiendi et al., OPAL-coll., preprint hep-ex/0001056
\bibitem{lit92}L. S. Littenberg, R. E. Shrock, \Journal{\PRL}{68}{443}{1992}
\bibitem{hal76} A. Halprin, P. Minkowski, H. Primakoff, S. P. Rosen,  
\Journal{\PRD}{13}{2567}{1976}, J. N. Ng, A. N. Kamal, \Journal{\PRD}{18}{3412}{1978},
J. Abad, J. G. Esteve, A. F. Pachero, \Journal{\PRD}{30}{1488}{1984}
\bibitem{cha68} C. Chang et al., \Journal{\PRL}{20}{510}{1968}
\bibitem{nis99}H. Nishiura, K. Matsuda, T. Fukuyama, \Journal{\MPL} {A 14}{433}{1999}
\bibitem{ma99} H. Ma, E865-coll., private communication
\bibitem{ket00} S. Kettell, preprint hep-ex/0002011
\bibitem{kod95} K. Kodama et al., E653-coll., \Journal{\PLB}{345}{85}{1995}
\bibitem{mis94} J. H. Missimer, R. N. Mohapatra, N. C. Mukhopadhyay, \Journal{\PRD}{50}{2067}{1994}
\bibitem{wei90} A. Weir et al., Mark2-coll. \Journal{\PRD}{41}{1384}{1990}
\end{thebibliography}
\end{document}